


\font\mayusc=cmcsc10 


      \font \ninebf                 = cmbx9
      \font \ninei                  = cmmi9
      \font \nineit                 = cmti9
      \font \ninerm                 = cmr9
      \font \ninesans               = cmss10 at 9pt
      \font \ninesl                 = cmsl9
      \font \ninesy                 = cmsy9
      \font \ninett                 = cmtt9
      \font \fivesans               = cmss10 at 5pt
						\font \sevensans              = cmss10 at 7pt
      \font \sixbf                  = cmbx6
      \font \sixi                   = cmmi6
      \font \sixrm                  = cmr6
						\font \sixsans                = cmss10 at 6pt
      \font \sixsy                  = cmsy6
      \font \tams                   = cmmib10
      \font \tamss                  = cmmib10 scaled 700
						\font \tensans                = cmss10
    
      \skewchar\ninei='177 \skewchar\sixi='177
      \skewchar\ninesy='60 \skewchar\sixsy='60
      \hyphenchar\ninett=-1
      \def\newline{\hfil\break}%
      \catcode`@=11
      \def\folio{\ifnum\pageno<\z@
      \uppercase\expandafter{\romannumeral-\pageno}%
      \else\number\pageno \fi}
      \catcode`@=12 

      \newfam\sansfam
      \textfont\sansfam=\tensans\scriptfont\sansfam=\sevensans
      \scriptscriptfont\sansfam=\fivesans
      \def\sans{\fam\sansfam\tensans}


      \def\petit{\def\rm{\fam0\ninerm}%
      \textfont0=\ninerm \scriptfont0=
\sixrm \scriptscriptfont0=\fiverm
       \textfont1=\ninei \scriptfont1=
\sixi \scriptscriptfont1=\fivei
       \textfont2=\ninesy \scriptfont2=
\sixsy \scriptscriptfont2=\fivesy
       \def\it{\fam\itfam\nineit}%
       \textfont\itfam=\nineit
       \def\sl{\fam\slfam\ninesl}%
       \textfont\slfam=\ninesl
       \def\bf{\fam\bffam\ninebf}%
       \textfont\bffam=\ninebf \scriptfont\bffam=\sixbf
       \scriptscriptfont\bffam=\fivebf
       \def\sans{\fam\sansfam\ninesans}%
       \textfont\sansfam=\ninesans \scriptfont\sansfam=\sixsans
       \scriptscriptfont\sansfam=\fivesans
       \def\tt{\fam\ttfam\ninett}%
       \textfont\ttfam=\ninett
       \normalbaselineskip=11pt
       \setbox\strutbox=\hbox{\vrule height7pt depth2pt width0pt}%
       \normalbaselines\rm


      \def\bvec##1{{\textfont1=\tbms\scriptfont1=\tbmss
      \textfont0=\ninebf\scriptfont0=\sixbf
      \mathchoice{\hbox{$\displaystyle##1$}}{\hbox{$\textstyle##1$}}
      {\hbox{$\scriptstyle##1$}}{\hbox{$\scriptscriptstyle##1$}}}}}


\def\imag{\mathop{\rm Im}}

.

					\mathchardef\gammav="0100
     \mathchardef\deltav="0101
     \mathchardef\thetav="0102
     \mathchardef\lambdav="0103
     \mathchardef\xiv="0104
     \mathchardef\piv="0105
     \mathchardef\sigmav="0106
     \mathchardef\upsilonv="0107
     \mathchardef\phiv="0108
     \mathchardef\psiv="0109
     \mathchardef\omegav="010A


					\mathchardef\Gammav="0100
     \mathchardef\Deltav="0101
     \mathchardef\Thetav="0102
     \mathchardef\Lambdav="0103
     \mathchardef\Xiv="0104
     \mathchardef\Piv="0105
     \mathchardef\Sigmav="0106
     \mathchardef\Upsilonv="0107
     \mathchardef\Phiv="0108
     \mathchardef\Psiv="0109
     \mathchardef\Omegav="010A



\font\grbfivefm=cmbx5
\font\grbsevenfm=cmbx7
\font\grbtenfm=cmbx10 
\newfam\grbfam
\textfont\grbfam=\grbtenfm
\scriptfont\grbfam=\grbsevenfm
\scriptscriptfont\grbfam=\grbfivefm

\font\calbfivefm=cmbsy10 at 5pt
\font\calbsevenfm=cmbsy10 at 7pt
\font\calbtenfm=cmbsy10 
\newfam\calbfam
\textfont\calbfam=\calbtenfm
\scriptfont\calbfam=\calbsevenfm
\scriptscriptfont\calbfam=\calbfivefm



      \def\bvec#1{{\textfont1=\tams\scriptfont1=\tamss
      \textfont0=\tenbf\scriptfont0=\sevenbf
      \mathchoice{\hbox{$\displaystyle#1$}}{\hbox{$\textstyle#1$}}
      {\hbox{$\scriptstyle#1$}}{\hbox{$\scriptscriptstyle#1$}}}}



\def\pmbf#1{\leavevmode\setbox0=\hbox{#1}%
\kern-.02em\copy0\kern-\wd0
\kern.04em\copy0\kern-\wd0
\kern-.02em\copy0\kern-\wd0
\kern-.03em\copy0\kern-\wd0
\kern.06em\box0 }



						\def\monthname{%
   			\ifcase\month
      \or Jan\or Feb\or Mar\or Apr\or May\or Jun%
      \or Jul\or Aug\or Sep\or Oct\or Nov\or Dec%
   			\fi
							}%
					\def\timestring{\begingroup
   		\count0 = \time
   		\divide\count0 by 60
   		\count2 = \count0   
   		\count4 = \time
   		\multiply\count0 by 60
   		\advance\count4 by -\count0   
   		\ifnum\count4<10
     \toks1 = {0}%
   		\else
     \toks1 = {}%
   		\fi
   		\ifnum\count2<12
      \toks0 = {a.m.}%
   		\else
      \toks0 = {p.m.}%
      \advance\count2 by -12
   		\fi
   		\ifnum\count2=0
      \count2 = 12
   		\fi
   		\number\count2:\the\toks1 \number\count4 \thinspace \the\toks0
					\endgroup}%

				\def\timestamp{\number\day\space
\monthname\space\number\year\quad\timestring}%
				\newskip\abovelistskip      \abovelistskip = .5\baselineskip 
				\newskip\interitemskip      \interitemskip = 0pt
				\newskip\belowlistskip      \belowlistskip = .5\baselineskip
				\newdimen\listleftindent    \listleftindent = 0pt
				\newdimen\listrightindent   \listrightindent = 0pt

				%


\def\petit{\def\rm{\fam0\ninerm}%
\textfont0=\ninerm \scriptfont0=\sixrm \scriptscriptfont0=\fiverm
\textfont1=\ninei \scriptfont1=\sixi \scriptscriptfont1=\fivei
\textfont2=\ninesy \scriptfont2=\sixsy \scriptscriptfont2=\fivesy
       \def\it{\fam\itfam\nineit}%
       \textfont\itfam=\nineit
       \def\sl{\fam\slfam\ninesl}%
       \textfont\slfam=\ninesl
       \def\bf{\fam\bffam\ninebf}%
       \textfont\bffam=\ninebf \scriptfont\bffam=\sixbf
       \scriptscriptfont\bffam=\fivebf
       \def\sans{\fam\sansfam\ninesans}%
       \textfont\sansfam=\ninesans \scriptfont\sansfam=\sixsans
       \scriptscriptfont\sansfam=\fivesans
       \def\tt{\fam\ttfam\ninett}%
       \textfont\ttfam=\ninett
       \normalbaselineskip=11pt
       \setbox\strutbox=\hbox{\vrule height7pt depth2pt width0pt}%
       \normalbaselines\rm
      \def\vec##1{{\textfont1=\tbms\scriptfont1=\tbmss
      \textfont0=\ninebf\scriptfont0=\sixbf
      \mathchoice{\hbox{$\displaystyle##1$}}{\hbox{$\textstyle##1$}}
      {\hbox{$\scriptstyle##1$}}{\hbox{$\scriptscriptstyle##1$}}}}}

      \def\footnoterule{\kern-3pt\hrule width 2cm\kern2.6pt}
      \newdimen\oldparindent\oldparindent=1.5em
      \parindent=1.5em
 
\newcount\footcount \footcount=0
\def\advftncnt{\advance\footcount by1\global\footcount=\footcount}
      \def\fnote#1{\advftncnt$^{\the\footcount}$\begingroup\petit
      \parfillskip=0pt plus 1fil
      \def\textindent##1{\hangindent0.5\oldparindent\noindent\hbox
      to0.5\oldparindent{##1\hss}\ignorespaces}%
 \vfootnote{$^{\the\footcount}$}
{#1\nullbox{0mm}{2mm}{0mm}\vskip-9.69pt}\endgroup}


      \def\item#1{\par\noindent
      \hangindent6.5 mm\hangafter=0
      \llap{#1\enspace}\ignorespaces}
      
      \def\leaderfill{\kern0.5em\leaders
\hbox to 0.5em{\hss.\hss}\hfill\kern
      0.5em}
						\def\hb{\hfill\break}

    \def\centerrule#1{\centerline{\kern#1\hrulefill\kern#1}}


      \def\boxit#1{\vbox{\hrule\hbox{\vrule\kern3pt
						\vbox{\kern3pt#1\kern3pt}\kern3pt\vrule}\hrule}}

      \def\tightboxit#1{\vbox{\hrule\hbox{\vrule
						\vbox{#1}\vrule}\hrule}}

      \def\looseboxit#1{\vbox{\hrule\hbox{\vrule\kern5pt
						\vbox{\kern5pt#1\kern5pt}\kern5pt\vrule}\hrule}}

      \def\youboxit#1#2{\vbox{\hrule\hbox{\vrule\kern#2
						\vbox{\kern#2#1\kern#2}\kern#2\vrule}\hrule}}




			\def\whitetile#1#2#3{\setbox0=\null
			\ht0=#1 \dp0=#2\wd0=#3 \setbox1=
\hbox{\tightboxit{\box0}}\lower#2\box1}

			\def\nullbox#1#2#3{\setbox0=\null
			\ht0=#1 \dp0=#2\wd0=#3\box0}




\def\equ{\leavevmode Eq.}

\def\equn#1{\ifmmode \eqno{\rm #1}\else \equ~#1\fi}



\def\tev{\ifmmode \mathop{\rm TeV}\nolimits\else {\rm TeV}\fi}
\def\gev{\ifmmode \mathop{\rm GeV}\nolimits\else {\rm GeV}\fi}
\def\mev{\ifmmode \mathop{\rm MeV}\nolimits\else {\rm MeV}\fi}
\def\kev{\ifmmode \mathop{\rm keV}\nolimits\else {\rm keV}\fi}
\def\ev{\ifmmode \mathop{\rm eV}\nolimits\else {\rm eV}\fi}

\def\chidof{\ifmmode
\mathop\chi^2/{\rm d.o.f.}\else $\chi^2/{\rm d.o.f.}\null$\fi}

\def\msbar{\ifmmode
\mathop{\overline{\rm MS}}\else$\overline{\rm MS}$\null\fi}


\def\physmatex{P\kern-.14em\lower.5ex\hbox{\sevenrm H}ys
\kern -.35em \raise .6ex \hbox{{\sevenrm M}a}\kern -.15em
 T\kern-.1667em\lower.5ex\hbox{E}\kern-.125emX\null}%

\def\ref#1{$^{[#1]}$\relax}

\def\prajnyp#1#2#3#4#5{
\frenchspacing{\mayusc #1}, {\sl#2}, {\bf #3}, {#5} {(#4)}}





\def\simeqsub{\mathop{\simeq}\limits}








\def\ddal{\mathop{\vrule height 7pt depth0.2pt
\hbox{\vrule height 0.5pt depth0.2pt width 6.2pt}
\vrule height 7pt depth0.2pt width0.8pt
\kern-7.4pt\hbox{\vrule height 7pt depth-6.7pt width 7.pt}}}
\def\sdal{\mathop{\kern0.1pt\vrule height 4.9pt depth0.15pt
\hbox{\vrule height 0.3pt depth0.15pt width 4.6pt}
\vrule height 4.9pt depth0.15pt width0.7pt
\kern-5.7pt\hbox{\vrule height 4.9pt depth-4.7pt width 5.3pt}}}
\def\ssdal{\mathop{\kern0.1pt\vrule height 3.8pt depth0.1pt width0.2pt
\hbox{\vrule height 0.3pt depth0.1pt width 3.6pt}
\vrule height 3.8pt depth0.1pt width0.5pt
\kern-4.4pt\hbox{\vrule height 4pt depth-3.9pt width 4.2pt}}}




\mathchardef\lap='0001


\def\lsim{\mathop{\setbox0=\hbox{$\displaystyle 
\raise2.2pt\hbox{$\;<$}\kern-7.7pt\lower2.6pt\hbox{$\sim$}\;$}
\box0}}
\def\gsim{\mathop{\setbox0=\hbox{$\displaystyle 
\raise2.2pt\hbox{$\;>$}\kern-7.7pt\lower2.6pt\hbox{$\sim$}\;$}
\box0}}

\def\gsimsub#1{\mathord{\vtop to0pt{\ialign{##\crcr
$\hfil{{\mathop{\setbox0=\hbox{$\displaystyle 
\raise2.2pt\hbox{$\;>$}\kern-7.7pt\lower2.6pt\hbox{$\sim$}\;$}
\box0}}}\hfil$\crcr\noalign{\kern1.5pt\nointerlineskip}
$\hfil\scriptstyle{#1}{}\kern1.5pt\hfil$\crcr}\vss}}}

\def\lsimsub#1{\mathord{\vtop to0pt{\ialign{##\crcr
$\hfil\displaystyle{\mathop{\setbox0=\hbox{$\displaystyle 
\raise2.2pt\hbox{$\;<$}\kern-7.7pt\lower2.6pt\hbox{$\sim$}\;$}
\box0}}
\def\gsim{\mathop{\setbox0=\hbox{$\displaystyle 
\raise2.2pt\hbox{$\;>$}\kern-7.7pt\lower2.6pt\hbox{$\sim$}\;$}
\box0}}\hfil$\crcr\noalign{\kern1.5pt\nointerlineskip}
$\hfil\scriptstyle{#1}{}\kern1.5pt\hfil$\crcr}\vss}}}

\def\ii{{\rm i}}
\def\dd{{\rm d}}

\def\ee{{\rm e}}
\def\gammae{\gamma_{\rm E}}




\def\frac#1#2{{#1\over#2}}
\def\dfrac#1#2{{\displaystyle{#1\over#2}}}
\def\tfrac#1#2{{\textstyle{#1\over#2}}}
\def\ffrac#1#2{\leavevmode
   \kern.1em \raise .5ex \hbox{\the\scriptfont0 #1}%
   \kern-.1em $/$%
   \kern-.15em \lower .25ex \hbox{\the\scriptfont0 #2}%
}%



\def\brochureb#1#2#3{\pageno#3
\headline={\ifodd\pageno{\rheadline}
\else\lheadline\fi}
\def\rheadline{\hfil -{#2}-\hfil}
\def\lheadline{\hfil-{#1}-\hfil}
\footline={\hss -- \number\pageno\ --\hss}
\voffset=2\baselineskip}

\def\nada{\phantom{M}\kern-1em}
\def\brochureendcover#1{\vfill\eject\pageno=1{\nada#1}\vfill\eject}





\def\chapterb#1#2#3{\pageno#3
\headline={\ifodd\pageno{\ifnum\pageno=#3\hfil\else\rheadline\fi}
\else\lheadline\fi}
\def\rheadline{\hfil -{#2}-\hfil}
\def\lheadline{\hfil-{#1}-\hfil}
\footline={\hss -- \number\pageno\ --\hss}
\voffset=2\baselineskip}


\def\bookendchapter{\ifodd\pageno
 \vfill\eject\footline={\hfill}\headline={\hfill}\null \vfill\eject
 \else\vfill\eject \fi}

\def\obookendchapter{\ifodd\pageno\vfill\eject
 \else\vfill\eject\null \vfill\eject\fi}


\def\booksection#1{
\setbox0=\vbox{\hsize=0.85\hsize\tolerance=500\raggedright\hfuzz=6mm
\noindent{\medfib #1}\medskip}\goodbreak\vskip0.6cm\box0
\nobreak
\noindent}
\def\booksubsection#1{
\setbox0=\vbox{\hsize=0.85\hsize\tolerance=400\raggedright\hfuzz=4mm
\noindent{\fib #1}\smallskip}\goodbreak\vskip0.45cm\box0
\nobreak
\noindent}








\def\abstracttype#1{
\hsize0.7\hsize\tolerance=800\hfuzz=0.5mm \noindent{\fib #1}\par
\medskip\petit}


\def\hb{\hfill\break}


\font\twelverm=cmr12 
\font\smallsc=cmcsc10 at 9pt 
\font\fib=cmfib8
\font\medfib=cmfib8 at 9pt


\font\sc=cmcsc10 

\font\addressfont=cmbxti10 at 9pt


\catcode`@=11 

\newdimen\pagewidth \newdimen\pageheight \newdimen\ruleht
 \maxdepth=2.2pt  \parindent=3pc
\pagewidth=\hsize \pageheight=\vsize \ruleht=.4pt
\abovedisplayskip=6pt plus 3pt minus 1pt
\belowdisplayskip=6pt plus 3pt minus 1pt
\abovedisplayshortskip=0pt plus 3pt
\belowdisplayshortskip=4pt plus 3pt

\newinsert\margin
\dimen\margin=\maxdimen




\newdimen\paperheight \paperheight = \vsize
\def\topmargin{\afterassignment\@finishtopmargin \dimen0}%
\def\@finishtopmargin{%
  \dimen2 = \voffset		
  \voffset = \dimen0 \advance\voffset by -1in
  \advance\dimen2 by -\voffset	
  \advance\vsize by \dimen2	
}%
\def\advancetopmargin{%
  \dimen0 = 0pt \afterassignment\@finishadvancetopmargin \advance\dimen0
}%
\def\@finishadvancetopmargin{%
  \advance\voffset by \dimen0
  \advance\vsize by -\dimen0
}%
\def\bottommargin{\afterassignment\@finishbottommargin \dimen0}%
\def\@finishbottommargin{%
  \@computebottommargin		
  \advance\dimen2 by -\dimen0	
  \advance\vsize by \dimen2	
}%
\def\advancebottommargin{%
  \dimen0 = 0pt\afterassignment\@finishadvancebottommargin \advance\dimen0
}%
\def\@finishadvancebottommargin{%
  \advance\vsize by -\dimen0
}%
\def\@computebottommargin{%
  \dimen2 = \paperheight	
  \advance\dimen2 by -\vsize	
  \advance\dimen2 by -\voffset	
  \advance\dimen2 by -1in	
}%
\newdimen\paperwidth \paperwidth = \hsize
\def\leftmargin{\afterassignment\@finishleftmargin \dimen0}%
\def\@finishleftmargin{%
  \dimen2 = \hoffset		
  \hoffset = \dimen0 \advance\hoffset by -1in
  \advance\dimen2 by -\hoffset	
  \advance\hsize by \dimen2	
}%
\def\advanceleftmargin{%
  \dimen0 = 0pt \afterassignment\@finishadvanceleftmargin \advance\dimen0
}%
\def\@finishadvanceleftmargin{%
  \advance\hoffset by \dimen0
  \advance\hsize by -\dimen0
}%
\def\rightmargin{\afterassignment\@finishrightmargin \dimen0}%
\def\@finishrightmargin{%
  \@computerightmargin		
  \advance\dimen2 by -\dimen0	
  \advance\hsize by \dimen2	
}%
\def\advancerightmargin{%
  \dimen0 = 0pt \afterassignment\@finishadvancerightmargin \advance\dimen0
}%
\def\@finishadvancerightmargin{%
  \advance\hsize by -\dimen0
}%
\def\@computerightmargin{%
  \dimen2 = \paperwidth		
  \advance\dimen2 by -\hsize	
  \advance\dimen2 by -\hoffset	
  \advance\dimen2 by -1in	
}%

\def\onepageout#1{\shipout\vbox{ 
    \offinterlineskip 
    \vbox to 3pc{ 
      \iftitle 
        \global\titlefalse 
        \setcornerrules 
      \else\ifodd\pageno \rightheadline\else\leftheadline\fi\fi
      \vfill} 
    \vbox to \pageheight{
      \ifvoid\margin\else 
        \rlap{\kern31pc\vbox to\z@{\kern4pt\box\margin \vss}}\fi
      #1 
      \ifvoid\footins\else 
        \vskip\skip\footins \kern-3pt
        \hrule height\ruleht width\pagewidth \kern-\ruleht \kern3pt
        \unvbox\footins\fi
      \boxmaxdepth=\maxdepth
      } 
    }
  \advancepageno}

\def\setcornerrules{\hbox to \pagewidth{\vrule width 1pc height\ruleht
    \hfil \vrule width 1pc}
  \hbox to \pagewidth{\llap{\sevenrm(page \folio)\kern1pc}%
    \vrule height1pc width\ruleht depth\z@
    \hfil \vrule width\ruleht depth\z@}}
\newbox\partialpage




\hyphenation{ha-ya-ka-wa acha-sov}
\def\ty{\hbox{TY-I}}


\input epsf.sty
\raggedbottom
\footline={\hfill}
\rightline{Revised \timestamp}
\smallskip
\rightline{July, 26, 2001}
\rightline{FTUAM  01-15}
\rightline{(hep-ph/0107318)}
\bigskip
\hrule height .3mm
\vskip.6cm
\centerline{{\twelverm Calculation of $\bar{\alpha}_{\rm Q.E.D.}$ 
on the $Z$}\footnote{(*)}{Work supported in part by CICYT, Spain}}
\medskip
\centerrule{.7cm}
\vskip1cm

\setbox9=\vbox{\hsize65mm {\noindent\fib 
J. F. de Troc\'oniz and F. J. 
Yndur\'ain} 
\vskip .1cm
\noindent{\addressfont Departamento de F\'{\i}sica Te\'orica, C-XI,\hb
 Universidad Aut\'onoma de Madrid,\hb
 Canto Blanco,\hb
E-28049, Madrid, Spain.}\hb}
\smallskip
\centerline{\box9}
\bigskip
\setbox0=\vbox{\abstracttype{Abstract} 
We perform a new, detailed calculation of the hadronic contributions 
to the running electromagnetic coupling, $\bar{\alpha}$, defined on the $Z$ particle (91 \gev).
We find for the hadronic 
contribution,  including radiative corrections,
$$10^5\times \deltav_{\rm had.}\alpha(M_Z^2)=
2740\pm12,$$
or, excluding the top quark contribution,
$$10^5\times \deltav_{\rm had.}\alpha^{(5)}(M_Z^2)=
2747\pm12.$$

Adding the pure QED corrections we get a  value for the running 
electromagnetic coupling of
$$\bar{\alpha}_{\rm Q.E.D.}(M_Z^2)=
{{1}\over{128.965\pm0.017}}.$$
}
\centerline{\box0}
\brochureendcover{Typeset with \physmatex}
\pageno=1
\brochureb{\smallsc j. f. de troc\'oniz and f. j.  yndur\'ain}{\smallsc 
calculation of $\bar{\alpha}_{\rm Q.E.D.}$ on the $Z$}{1}

\booksection{1. Introduction}
In a recent paper,\ref{1} hereafter to be referred to as TY-I, 
we have evaluated the hadronic contributions to 
the anomalous magnetic moment of the muon; 
specifically, a very precise determination of the 
piece involving the photon vacuum polarization function 
was given there. 
With a simple change of integration kernel (see below) 
this analysis can be extended to evaluate the hadronic contribution to the 
QED running coupling, $\bar{\alpha}_{\rm Q.E.D.}(t)$, in 
particular for $t=M^2_Z$; an important quantity that enters 
into precision evaluations of electroweak observables. 
We will find that we can produce a substantial improvement over previous 
determinations due to our use of complete and correct analyticity and 
unitarity\ref{2} properties (at low energy), and the high quality of recent 
Novosibirsk, LEP, and Beijing data.

The running coupling constant may be written as
$$\bar{\alpha}_{\rm Q.E.D.}(t)=\dfrac{e^2/4\pi}{1+\hat{\piv}(t)},
\quad \hat{\piv}(t)\equiv e^2 \piv_{\rm ren}(t)\equn{(1.1)}$$
where $e$ is the electron charge and $\piv(t)_{\rm ren}$ is the 1PI (one particle irreducible) 
vacuum polarization function, 
renormalized at $t=0$. 
To lowest order we can write the shift in $\alpha$ as
$$\bar{\alpha}_{\rm Q.E.D.}(t)=\left\{1+\deltav \alpha(t)\right\}\dfrac{e^2}{4\pi},
\quad \deltav \alpha(t)=-e^2\piv_{\rm ren}(t).$$ 
In fact, we will evaluate $\piv_{\rm ren}(t)$ including the first 
radiative corrections, so the full (1.1) has to be used 
to find the effective coupling. However, we will follow current usage and will write, 
for the hadronic contributions $\hat{\piv}_{\rm h}\equiv e^2\piv^{\rm had}_{\rm ren}$,
$$\deltav_{\rm had}\alpha\equiv-\hat{\piv}_{\rm h}=-e^2\piv^{\rm had}_{\rm ren}$$
or, distinguishing between lowest order (index 0) and next order (index 1),
$$\deltav_{\rm had}^{(0)}\alpha\equiv-\hat{\piv}_{\rm h}^{(0)}=-e^2\piv^{\rm had; (0)}_{\rm ren},\quad
\deltav_{\rm had}^{(1)}\alpha\equiv-\hat{\piv}_{\rm h}^{(1)}=-e^2\piv^{\rm had; (1)}_{\rm ren}.$$ 

By using a dispersion relation one can write this  hadronic 
contribution at energy squared $t$,  $\hat{\piv}_h(t)$, as 
$$-\hat{\piv}_{\rm h}(t)\equiv -e^2\piv_{\rm ren}^{\rm had}(t)=-\dfrac{t\alpha}{3\pi}
\int_{4m^2_\pi}^\infty 
\dd s\;\dfrac{R(s)}{s(s-t)},\equn{(1.2a)}$$
with
$$R(s)=\dfrac{\sigma(e^+e^-\to{\rm hadrons};\,s)}{\sigma^{(0)}(e^+e^-\to\mu^+\mu^-;\,s)},
\quad \sigma^{(0)}(e^+e^-\to\mu^+\mu^-;\,s)\equiv\dfrac{4\pi\alpha^2}{3s},\equn{(1.2b)}$$
and the integral in (1.2a) has to be understood as a principal part integral.
This is similar to the Brodsky--de~Rafael expression for the hadronic contribution to the 
muon magnetic moment anomaly,

$$\eqalign{a(\hbox{h.v.p.})=&\int_{4m^2_\pi}^\infty \dd s\,K(s) R(s),\cr
K(s)=&\dfrac{\alpha^2}{3\pi^2s}\hat{K}(s);\quad \hat{K}(s)=
\int_0^1\dd x\,\dfrac{x^2(1-x)}{x^2+(1-x)s/m^2_\mu}.\cr}
$$
Therefore, we can carry over all the work from \ty\ with the simple replacement   
$$K(s)\to-\dfrac{t\alpha}{3\pi}\dfrac{1}{s(s-t)}.$$
For the coupling at the $Z$ we will take $t=M_Z^2$.
Because of this similarity with the $g-2$ calculation,
 we will dispense with many discussions or details; they may be found in 
TY-I. Indeed, the present paper should be considered as a sequel to 
the former one.
 
After the corresponding evaluations we find, to next to leading order in $\alpha$,
$$10^5\times\deltav_{\rm had}\alpha(M_Z^2)=-10^5\times\left[\hat{\piv}^{(0)}_{\rm h}(M_Z^2)
+\hat{\piv}^{(1)}_{\rm h}(M_Z^2)\right]=
2740\pm12,\equn{(1.3)}$$
or, excluding the top quark contribution,
$$10^5\times\deltav_{\rm had}\alpha^{(5)}(M_Z^2)=
2747\pm12.\equn{(1.4)}$$
Adding the known pure QED corrections,
the running QED coupling, in the momentum scheme is
$$\bar{\alpha}_{\rm Q.E.D.}(M_Z^2)=
\dfrac{1}{128.965\pm0.017}.
\equn{(1.5)}$$

\booksection{2. Contributions to the lowest order $-\hat{\piv}^{(0)}_{\rm h}$ in
 the energy range from threshold to $2\;\gev^2$}
\vskip-0.3cm
\booksubsection{2.1. The region $s\leq 1.2\;\gev^2$}
To zero order in the e.m. interactions we can write  $-\hat{\piv}^{(0)}_{\rm h}$ 
as a sum of contributions of different intermediate states in 
various energy slices. 
We start with the $2\pi$ states for $s\leq 1.2\;\gev^2$. This we 
will subdivide in turn into two pieces: 
from threshold, $4m^2_\pi$, to $0.8\;\gev^2$, and the higher energy piece.
\medskip
\noindent 2.1.1. {\sl The region below $0.8\,\gev^2$}
\smallskip
\noindent  
We can express $R^{(0)}$ in terms of the pion form factor, $F_\pi$:
$$R^{(0)}(s)=\dfrac{1}{4}\left(1-\dfrac{4m^2_\pi}{s}\right)^{3/2} 
|F_\pi(s)|^2,\equn{(2.1)}$$
where by $m_\pi$ we understand the charged pion mass. 
We can also relate $F_\pi$ to the decay $\tau^+\to\bar{\nu}_\tau\pi^+\pi^0$. 
Consider the correlator
$$\Piv^V_{\mu\nu}=
\ii\int\dd^4x\,\ee^{\ii p\cdot x}\langle0|{\rm T}V^+_\mu(x)V_\nu(0)|0\rangle=
\left(-p^2g_{\mu\nu}+p_\mu p_\nu\right)\Piv^V(s)+p_\mu p_\nu \piv^{S}(s),\; s=p^2;
\equn{(2.2a)}$$
with $V_\mu$ the weak vector current.
Then neglecting isospin breaking (except for 
the phase space factor) we have at low $s$,
$$v_1(s)\equiv2\pi\imag \Piv^{V} =\tfrac{1}{12}
\left\{\left[1-\dfrac{(m_{\pi^+}-m_{\pi^0})^2}{s}\right]
\left[1-\dfrac{(m_{\pi^+}+m_{\pi^0})^2}{s}\right] \right\}^{3/2}|F_\pi(s)|^2,
\equn{(2.2b)}$$
and, on the other hand, $v_1$ may be obtained from 
the experimental measurements of the decay  $\tau^+\to\bar{\nu}_\tau\pi^+\pi^0$.

To obtain $F_\pi(s)$ we will fit the 
recent Novosibirsk data\ref{3} on $e^+e^-\to\pi^+\pi^-$ and the 
tau decay data of Aleph and Opal.\ref{3} 
We will take into account, at least partially, isospin breaking effects 
by allowing different masses and widths for the $\rho^0$, $\rho^+$ resonances. 
Moreover, and to get a good grip in the low energy region where data are inexistent or very poor, 
we also fit $F_\pi(s)$ at spacelike $s$.\ref{3} 
This is possible in our approach because we use an expression for $F_\pi$ that takes fully into account 
its analyticity properties.\ref{2} To be precise, we use that the phase of 
$F_\pi(s)$ is equal to that of 
$\pi\pi$ scattering, in the elastic region, and then the Omn\`es-Muskhelishvili method. 
We write
$$F_\pi(s)=G(s)J(s).
\equn{(2.3a)}$$
Here $J$ is expressed in terms of the P-wave $\pi\pi$ phase shift, $\delta_1^1$, as 
$$J(s)=\ee^{1-\delta_1^1(s_0)/\pi}
\left(1-\dfrac{s}{s_0}\right)^{[1-\delta_1^1(s_0)/\pi]s_0/s}
\left(1-\dfrac{s}{s_0}\right)^{-1}
\exp\left\{\dfrac{s}{\pi}\int_{4m^2_\pi}^{s_0} \dd s'\;
\dfrac{\delta_1^1(s')}{s'(s'-s)}\right\}.
\equn{(2.3b)}$$
$s_0$ is the energy at which inelasticity starts becoming important (in practice, above the 
percent level); we will take $s_0=1.1\,\gev^2$ in actual calculations.

The exponential factor 
$$\exp\left\{\dfrac{s}{\pi}\int_{4m^2_\pi}^{s_0} \dd s'\;
\dfrac{\delta_1^1(s')}{s'(s'-s)}\right\}$$
in \equn{(2.3b)} guarantees that the phase of $J(s)$ is equal to $\delta^1_1(s)$ for $s\leq s_0$, 
hence equal also to the phase of  $F_\pi$. The rest is included so that $J$ is smooth at $s=s_0$, and 
has the behaviour $|J(s)|\sim 1/s$ at large energies. 

Because of this equality of the phase of $J(s)$ and the phase of $F_\pi(s)$ below $s=s_0$, it 
follows that $G(s)$ will be an analytic function also for 
$4m^2_\pi\leq s\leq s_0$, so in the whole $s$ plane except in a cut from 
$s=s_0$ to $+\infty$. If we now make the conformal transformation
$$z=\dfrac{\tfrac{1}{2}\sqrt{s_0}-\sqrt{s_0-s}}{\tfrac{1}{2}\sqrt{s_0}+\sqrt{s_0-s}}
\eqno{(2.4a)}$$
then, as a function of $z$, $G$ will be analytic in the unit disk and we can thus 
write a convergent Taylor series for it. 
Incorporating the condition $G(0)=1$, that follows from $F_\pi(0)=1$, 
and undoing the transformation, we have
$$G(s)=1+
c_1\left[\dfrac{\tfrac{1}{2}\sqrt{s_0}-\sqrt{s_0-s}}{\tfrac{1}{2}\sqrt{s_0}+\sqrt{s_0-s}}
+\tfrac{1}{3}\right]+
c_2\Bigg[\left(\dfrac{\tfrac{1}{2}\sqrt{s_0}-\sqrt{s_0-s}}{\tfrac{1}{2}\sqrt{s_0}+\sqrt{s_0-s}}\right)^2
-\tfrac{1}{9}\Bigg]+\cdots,
\eqno{(2.4b)}$$
$c_1,\,c_2,\,\dots$ free parameters. Actually, only two terms will be necessary to fit the data.

Next, to obtain $J$, and hence $F_\pi$, we need a parameterization of 
$\delta^1_1(s)$. 
We can use the well-known effective range theory to write
$$\cot \delta_1^1(s)= \dfrac{s^{1/2}}{2k^3}(m^2_\rho-s)\hat{\psi}(s),\quad k=\dfrac{\sqrt{s-4m^2_\pi}}{2};
\equn{(2.5)}$$
where we have extracted the zero corresponding to the rho resonance. 
Now the effective range function $\hat{\psi}(s)$ is analytic in the full $s$ plane
 except for a cut for $[-\infty,0]$ and the inelastic cut $[s_0,+\infty]$. 
We can profit from this analyticity by making again a conformal transformation
 into the unit circle, which is now given by
$$w=
\dfrac{\sqrt{s}-\sqrt{s_0-s}}{\sqrt{s}+\sqrt{s_0-s}}.
\equn{(2.6)}$$
We can therefore expand $\hat{\psi}$ in a convergent series\fnote{It is to 
be noted that this series, as well as that in terms of $z$ above, are 
quickly convergent in the region of interest for us here, which is mapped in 
segments contained in $[-0.57,0.24]$ inside the unit circles; see \ty\ for details.} of powers of $w$.
Undoing the transformation we then have
$$\delta_1^1(s)={\rm Arc\; cot}\left\{\dfrac{s^{1/2}}{2k^3}
(m^2_\rho-s)\left[b_0+b_1\dfrac{\sqrt{s}-\sqrt{s_0-s}}{\sqrt{s}+\sqrt{s_0-s}}+
\cdots\right]\right\}.
\equn{(2.7)}$$
We note that $b_0=\hbox{Const.}$, $b_{i\geq 1}=0$ 
would correspond to a pure Breit--Wigner shape for the rho. 
By allowing for more terms in the expansion we are 
taking into account the known distortions of the Breit--Wigner shape 
due to the influence of the left and the inelastic cuts of $\hat{\psi}$. 
For the actual fits, only $b_0,\,b_1$, and $m_\rho$ are needed as 
parameters.

The values of the parameters are obtained by fitting experimental data 
on $e^+e^-\to\pi^+\pi^-$, data  
on $\tau^+\to\bar{\nu}_{\tau}\pi^+\pi^0$ decay, and data on $F_\pi(s)$ at spacelike 
$s$ (ref.~3). We also include in the fit the value of the $\pi\pi$ 
P-wave scattering length, that we constrain at
$$a^1_1=(38\pm3)\times 10^{-3}\;m_\pi^{-3},\equn{(2.8)}$$
consistent with $\pi\pi$ scattering results as well as with current algebra 
calculations. For the free parameters of our fit we find 
$$
c_1=0.23\pm0.02,\;c _2=-0.15\pm0.03;\;b_0=1.062\pm0.005,\;b_1=0.25\pm0.04;\;
m_{\rho^0}=772.6\pm0.5\;\mev.
\equn{(2.9)}$$
 We also find, as byproduct of our fit, the $\rho^0$ width as 
well as the mass and the width of the $\rho^+$, the 
P-wave scattering length, and the mean square radius and second coefficient associated 
with the form factor of the  pion:
$$\gammav_{\rho^0}=147.4\pm0.8,\quad a^1_1=(41\pm2)\times10^{-3}m^{-3}_\pi,
\quad \langle r^2_\pi\rangle=0.435\pm0.002\;{\rm fm},\quad c_\pi=3.60\pm0.03\;\gev^{-4}
\equn{(2.10a)}$$
and
$$m_{\rho^+}=773.8\pm0.6\;\mev,\quad \gammav_{\rho^+}=147.3\pm0.9\,\mev.
\equn{(2.10b)}$$
The \chidof\ of the fit is $246/204$ with only 
statistical errors, but improves to $214/204$
when experimental systematic errors are included.
 
We have {\sl not} included in this fit the 
experimental $\pi\pi$ phase shifts 
(except for the scattering length), as they are known to suffer from 
uncertainties associated with the method of extraction: 
$\pi\pi$ scattering cannot be measured directly. 
However, we have checked that adding them would not alter 
substantially our fit or parameters. 
Details of this, and other aspects of the calculation, may be found in \ty, 
where also the results of separate fits to $e^+e^-$ and tau decay data are presented.

With the above parameterization of $F_\pi$ we can evaluate immediately  
the corresponding contribution to $\hat{\piv}$. We find, with self-explanatory notation,
$$-10^5\times \hat{\piv}^{(0)}_{\rm h}(M_Z^2;\;2\pi;\; s\leq 0.8\,\gev^2)=
307.6\pm2.2\pm2.9.\equn{(2.11a)}$$
The first error is statistical, the second a combination of 
systematic (taking into account the correlations among the various sets of 
experimental data), and theoretical ones.\fnote{If we had used 
data on $e^+e^-$, but not on $\tau$ decay, we would have obtained a 
slightly smaller number and a larger error: 
$-10^5\times \hat{\piv}^{(0)}_{\rm h}(M_Z^2;\;2\pi;\; s\leq 0.8\,\gev^2)=
306.5\pm4.0\pm4.3$. We will take (2.11a) to be our 
best result here.} (2.11a) 
includes the (small) effect of $\omega - \rho$ mixing, 
evaluated with the standard Gounnaris--Sakurai method 
as in \ty.

Errors included in this work are divided into statistical and systematic.
Evaluation of the statistical errors is standard: the fit procedure
(using the program MINUIT) provides the full error (correlation)
matrix at the $\chi^2$ minimum. This matrix is used when calculating the
corresponding integral for $\hat{\piv}$, therefore incorporating
automatically all the correlations among the various fit parameters.

In addition, for every energy region, we have considered the errors
that stem from experimental systematics, as well as those originating
from deficiencies of the theoretical analysis. The experimental systematics 
covers the errors given by the individual experiments included in the fits. 
Also, when conflicting sets of data exist, the calculation has been repeated, 
and the given systematic error bar enlarged to encompass all the possibilities.
In general, errors (considered as uncorrelated) have been added in quadrature.
The exceptions are explicitly discussed along the text.

\medskip
\noindent 2.1.2. {\sl The $\pi\pi$ contribution in the region $0.8\leq s\leq 1.2\,\gev^2$}
\smallskip
\noindent  
For the contribution in the region $0.8\leq s\leq1.2\,\gev^2$ we integrate numerically 
the experimental data,\ref{4} and get
$$-10^5\times \hat{\piv}^{(0)}_{\rm h}(M_Z^2;\;2\pi;\;0.8\leq\,s\leq1.2 \,\gev^2)=
27.3\pm0.3\pm0.5.\equn{(2.11b)}$$
With the result above,
$$-10^5\times \hat{\piv}^{(0)}_{\rm h}(M_Z^2;\;2\pi;\;4m^2_\pi\leq s\leq1.2 \,\gev^2)=
334.9\pm4.1\equn{(2.12)}$$
where both systematic errors (related to the same normalization uncertainty) have been added 
coherently.

\medskip
\noindent 2.1.2. {\sl The $3\pi$, $2K$, and other contributions in the region $ s\leq 1.2\,\gev^2$}
\smallskip
\noindent
For the $3\pi$ contribution 
we fit experimental data,\ref{4} with Breit--Wigner formulas (including the correct threshold 
behaviour) for the $\omega$, $\phi$ resonances, plus a constant. 
We have two sets of experimental data; the difference between the 
evaluations with each of them is included into the systematic error. 
The \chidof\ is $63/60$. The contribution to 
$\hat{\piv}^{(0)}_{\rm h}$ is,
$$-10^5\times \hat{\piv}^{(0)}_{\rm h}(M_Z^2;\;3\pi;\;9m^2_\pi\leq s\leq1.2 \,\gev^2)=
39.5\pm0.3\pm1.5.\equn{(2.13)}$$

The $2K$ states are treated in the same manner, fitting simultaneously 
$e^+e^-\to K_LK_S$ and $e^+e^-\to K^+K^-$ data\ref{4} with the same Breit--Wigner 
parameters for the $\phi$; the \chidof\ is $84/82$. For details we refer 
again to \ty.  
We get
$$-10^5\times \hat{\piv}^{(0)}_{\rm h}(M_Z^2;\;2K;\; s\leq1.2 \,\gev^2)=
41.6\pm0.2\pm1.3.\equn{(2.14)}$$

The contribution of $4\pi$ states is evaluated by numerical integration, with the 
trapezoid rule, of experimental data.\ref{5} 
The systematic error includes the (estimated) difference between evaluations based on different sets of 
experimental data. This gives the result,
$$-10^5\times \hat{\piv}^{(0)}_{\rm h}(M_Z^2;\;4\pi;\; s\leq1.2 \,\gev^2)=2.6\pm0.7\equn{(2.15)}$$

Finally, $5\pi,\;6\pi,\;\eta\pi\pi,\,\dots$ states contribute
 $(0.3\pm0.2)\times 10^{-5}$ in this region.
If we add all the contributions with $s\leq 1.2\,\gev^2$ we find,
$$-10^5\times \hat{\piv}^{(0)}_{\rm h}(M_Z^2;\;s\leq1.2 \,\gev^2)=
418.9\pm4.6\equn{(2.16)}$$
\medskip
\booksubsection{2.2. The energy range  $1.2\;\gev^2\leq s\leq 2\;\gev^2$}
We  have now  a 
numerical evaluation obtained from a  fit to inclusive $e^+e^-\to\hbox{hadrons}$  
experimental data:\ref{5}
$$-10^5\times \hat{\piv}^{(0)}_{\rm h}(M_Z^2;\;1.2\,\gev^2\leq s\leq2 \,\gev^2)=
53.1\pm5.3.\equn{(2.17)}$$

\booksection{3. The lowest order $-\hat{\piv}^{(0)}_{\rm h}$ in
 the energy range above $2\;\gev^2$. The full $\hat{\piv}^{(0)}_{\rm h}(M_Z^2)$}
\vskip-0.3cm
\booksubsection{3.1. QCD calculations}
For the QCD calculations\fnote{See ref.~6 for 
the calculations of the various pieces.}
 we take the following approximation:  away from quark thresholds, and for $n_f$ massless quark 
flavours 
with charges $Q_f$, we write
$$\eqalign{R^{(0)}(t)=&\;3\sum_fQ_f^2\Bigg\{1+\dfrac{\alpha_s}{\pi}+
(1.986-0.115n_f)\left(\dfrac{\alpha_s}{\pi}\right)^2\cr
+&\;\Big[-6.64-1.20n_f-0.005n_f^2-1.240\dfrac{(\sum_f Q_f)^2}{3(\sum_f Q_f^2)}\Big]
\left(\dfrac{\alpha_s}{\pi}\right)^3\Bigg\}.\cr}
\equn{(3.1a)}$$
To this one adds mass and nonperturbative corrections. 
We take into account the quark mass effect for quarks with 
running mass $\bar{m}_i(s)$ which correct $R^{(0)}$ by the amount, for each quark, 
$$\eqalign{-3Q_i^2&\bar{m}^2_i(s)\left\{6+28\dfrac{\alpha_s}{\pi}
+(294.8-12.3n_f)\left(\dfrac{\alpha_s}{\pi}\right)^2\right\}s^{-1}\cr
+3Q_i^2&\,\dfrac{8\bar{m}^4_i(t)}{7}\left\{-\dfrac{6\pi}{\alpha_s}+
\tfrac{23}{4}+\left(\tfrac{2063}{24}-10\zeta(3)\right)\dfrac{\alpha_s}{\pi}\right\}s^{-2}.\cr}
\equn{(3.1b)}$$
Finally, for the condensates we add
$$\dfrac{2\pi}{3s^2}\left(1-\tfrac{11}{18}\dfrac{\alpha_s}{\pi}\right)\langle\alpha_s G^2\rangle
\sum_f Q^2_f
\equn{(3.1c)}$$
and
$$\dfrac{24\pi^2}{s^2}\left[1-\tfrac{23}{27}\dfrac{\alpha_s}{\pi}\right]m_i\langle\bar{\psi}_i\psi_i\rangle.
\equn{(3.1d)}$$
We neglect the condensates corresponding to heavy quarks ($c,\,b$) and 
express those for $u,\,d,\,s$ in terms of 
$f^2_\pi m^2_\pi$, $f^2_K m^2_K$ using the well-known PCAC relations. 
The condensate contributions are negligible above $s=3\gev^2$.

\equn{(3.1b)} will be used when $\bar{m}_i^2\ll s$. In practice this will mean that the 
contribution of the correction of order $\bar{m}_i^4/s^2$ is less than 
$10^{-5}$. Near the threshold for heavy quarks $c,\,b,\,t$, i.e., 
when $v_i^2(s)\ll 1$ (with $v_i(s)=(1-4m_i^2/s)^{1/2}$ 
the velocity of the quark) 
we use a nonrelativistic QCD calculation (see refs.~7 for details) 
in which the contribution of quark $i$ is
$$R_i^{\rm NR}=3Q^2_i\left[1+2c_0(s)\right]
\dfrac{3-v_i^2(s)}{2}\,\dfrac{\pi C_F\widetilde{\alpha}_s}{1-\ee^{-\pi C_F\widetilde{\alpha}_s/v_i}};
\equn{(3.2a)}$$
$$\eqalign{\widetilde{\alpha}_s(s)=&\left[1+
\dfrac{(93-10n_f)/36+\gammae\beta_0/2}{\pi}\alpha_s\right]\alpha_s(s),\cr
c_0(s)\simeqsub_{v_i\to0}&\;\dfrac{\beta_0\alpha_s}{4\pi}
\left\{\log\dfrac{s^{1/2}}{m_iC_F\widetilde{\alpha}_s}-1-2\gammae\right\}.\cr
}\equn{(3.2b)}$$
To this we add the leading nonperturbative correction
$$-\dfrac{2\pi\langle \alpha_s G^2\rangle}{192m_i^4v_i^6}$$
and consider the effective threshold to occur when  this 
overcomes the contribution (3.2a).

In the intermediate region between $v_i^2\ll1$ and $m_i^2\ll s$, we use the 
interpolation given by Schwinger\ref{8}
$$R_i^{\rm Schw.}=3Q^2_iv_i(s)\dfrac{3-v_i^2(s)}{2}\left\{1+
C_F\left[\dfrac{\pi}{3v_i}+
\dfrac{3+v_i}{4}\left(\dfrac{\pi}{2}-\dfrac{3}{4\pi}\right)\right]\alpha_s\right\}.
\equn{(3.3)}$$
Note, however, that Schwinger's interpolation {\sl cannot} be used for $v_i\to0$ as it 
underestimates $R_i$ by a factor of 2.

In the QCD calculations, the error labeled ``Cond." is found by 
inserting the variation obtained  
setting quark and gluon condensates to zero, and that labeled 
$\lambdav$ by varying the QCD parameter. Likewise, we label $m_i$ to the error 
obtained varying the mass $m_i$. If an error is not  given it will mean that it falls below the 
$10^{-5}$ level.

For the parameter $\lambdav$
we take the recent determinations\ref{9} 
that correspond to the value
$$\alpha_s(M_Z^2)=0.117\pm 0.003;$$
to be precise, we have taken (in \mev, and to four loops),
$$\lambdav(s\leq m_c^2)=373\pm80;\; \lambdav(m_c^2\leq s\leq m_b^2)=283\pm50;\;
\lambdav(m_b^2\leq s\leq m^2_t)=199\pm30;\; \lambdav(s\geq m_t)=126\pm 20.$$
For the gluon condensate we take $\langle\alpha_s G^2\rangle=0.07\,\gev^4$. 
Finally, for the running quark masses we take 
$$\bar{m}_s(1\,\gev)=0.188\;\gev;\quad \bar{m}_c(\bar{m}_c)=1.44\,\gev;
\quad \bar{m}_b(\bar{m}_b)=4.3\,\gev;\quad \bar{m}_t(\bar{m}_t)=174\,\gev,$$
and, for the pole masses,
$$m_c=1.867\pm0.20\,\gev;\quad m_b=5.022\pm0.060\,\gev;\quad m_t=174\pm5\,\gev.$$
For the $c,\,b$ masses, see refs.~10,11; for the $t$ quark, ref.~12.

\booksubsection{3.2. The regions away from quark thresholds}
At the lowest energy region we find
$$-10^5\times \hat{\piv}^{(0)}_{\rm h}(M_Z^2;\;2 \,\gev^2\leq s\leq3 \,\gev^2)=
71.1\pm0.5\;(\lambdav)\pm0.4\;({\rm Cond.)};\equn{(3.4)}$$
the justification of the applicability of QCD in this range is the agreement, within errors, of the 
QCD calculation with the $e^+e^-\to \hbox{hadrons}$ data, and with the 
more precise data coming from $\tau\to\nu_\tau+{\rm hadrons}$; 
this may be seen depicted in e.g. the plots of Aleph and Opal data in ref.~3 
(more details  may be found in \ty).

Apart from this region $2 \,\gev^2\leq s\leq 3\,\gev^2$, 
we can use the perturbative QCD formulas (3.1), (3.3) for the energy regions $s\geq 3\,\gev^2$ 
provided we stay away from heavy quark thresholds. 
We  will thus get, excluding the 
$J/\psi$, $\psi'$ resonances contributions (to be discussed below):
$$-10^5\times \hat{\piv}^{(0)}_{\rm h}(M_Z^2;\;3\,\gev^2 \leq s\leq3.7^2 \,\gev^2)=
259.1\pm1.5\,(\lambdav)\equn{(3.5a)}$$
(here the contribution of the error induced by the condensates is already negligible). 
Then,
$$-10^5\times \hat{\piv}^{(0)}_{\rm h}(M_Z^2;\;4.6^2\,\gev^2 \leq s\leq10.086^2 \,\gev^2)=
421.3\pm0.8\;(\lambdav).\equn{(3.5b)}$$

We will separate a region around $M^2_Z$, because we take the principal vale of the integral. 
We have thus,
$$\eqalign{
-10^5\times \hat{\piv}^{(0)}_{\rm h}(M_Z^2;\;11.2^2\,\gev^2 \leq s\leq 20^2\,\gev^2)=&\;
352.2\pm0.9;\cr
-10^5\times \hat{\piv}^{(0)}_{\rm h}(M_Z^2;\;20^2\,\gev^2 \leq s\leq(M_Z-3\,\gev)^2)=&\;
1668.9\pm0.9;\cr
-10^5\times \hat{\piv}^{(0)}_{\rm h}(M_Z^2;\;(M_Z-3\,\gev)^2\,\gev^2 \leq s\leq(M_Z+3\,\gev)^2)=&\;
29.2\pm0.5;\cr
-10^5\times \hat{\piv}^{(0)}_{\rm h}(M_Z^2;\;(M_Z+3\,\gev)^2\,\gev^2 \leq s\leq348^2 \,\gev^2)=&\;
-794.5\pm0.7.\cr
}\equn{(3.5c)}$$
All the errors are due to the variation of the parameter $\lambdav$. 
Finally,
$$\eqalign{-10^5\times \hat{\piv}^{(0)}_{\rm h}(M_Z^2;\;360^2\,\gev^2 \leq s\leq 400^2\,\gev^2)=&
\;-4.7\pm0.3\cr
-10^5\times \hat{\piv}^{(0)}_{\rm h}(M_Z^2;\;400^2\,\gev^2 \leq s\to\infty)=&\;-20.8\pm0.1.\cr}
\equn{(3.5d)}$$
In particular, the total top quark contribution above threshold 
($360^2\,\gev^2 \leq s\to\infty$) is -6.5.

We note that part of the ranges contain some of the narrow resonances ($\psi$, $\upsilonv$ and 
$T$ families). We will add their contributions individually later on. 
For the whole perturbative QCD contributions we have, adding \equn{(3.4)} to  \equn{(3.5)},
$$-10^5\times \hat{\piv}^{(0)}_{\rm h}(M_Z^2;\;2\,\gev^2 \leq s;\;\hbox{pQCD})=
1982\pm7.
\equn{(3.6)}$$
Note that we have added the errors {\sl linearly} as they stem from the 
same variation in the QCD parameter $\lambdav$.

\booksection{3.3. The thresholds regions}
We will make two types of calculations. In the first, we take experimental data 
(when possible, i.e., at the $\bar{c}c$ and $\bar{b}b$ thresholds); 
in the second, we take the contribution of the resonances lying below 
threshold from experiment, plus a background 
given by the contribution of the light quarks 
(evaluated with perturbative QCD, as above) and use nonrelativistic QCD to evaluate 
the contribution of the quarks whose threshold we are crossing. 
Of course, for the $\bar{t}t$ threshold this is all we have.
\medskip
\noindent3.3.1. {\sl   $\bar{c}c$: $J/\psi$, $\psi'$ and the continuum
 $3.7^2\;\gev^2\leq s\leq 4.6^2\;\gev^2$}
\smallskip
\noindent
We split this into the contribution of the $J/\psi$, $\psi'$, 
that we calculate in the n.w.a. (narrow width approximation), and the rest. For the first we have,
$$\eqalign{
10^{-5}\times (69.9\pm 4.5)&\quad [J/\psi]\cr
10^{-5}\times (23.6\pm2.1)&\quad [\psi']\cr}$$
\medskip
\goodbreak

For the remainder we have two possibilities: use a NRQCD calculation (see below) 
for the heavy quark, which gives
$$\eqalign{
10^{-5}\times (73.2\pm0.3\,(\lambdav))&\quad [uds];
\;\hbox{(QCD;  $3.7^2\;\gev^2\leq s\leq4.6^2\;\gev^2$)}\cr
10^{-5}\times (66\pm13\,(m_c))&\quad\hbox{[$\bar{c}c$]. (NRQCD)}\cr
\hbox{Sum}: 10^{-5}\times (139\pm13)}$$
\smallskip
\noindent{\sc Total:} $10^{-5}\times (233\pm14)\qquad$ (QCD+NRQCD).

Otherwise, we use experimental data above $3.7^2\;\gev^2$:

$$10^{-5}\times (111.8\pm0.6\pm5.5) \quad\hbox{(Exp., BES)}: 
3.7^2\;\gev^2\leq s\leq 4.6^2\;\gev^2$$
\smallskip
\noindent{\sc Total:} $10^{-5}\times (205.3\pm7.4)\qquad$ (Exp., BES)
\medskip
 (NRQCD) refers to the nonrelativistic QCD calculation with 
\equn{(3.2)}; see \ty\ and refs.~7 for details of this 
type of calculations. 
BES are the experimental data from ref.~13. The first error 
for them is the statistical, the second the systematic one. 

We give a few more details on the calculation with NRQCD, as it is 
the model for the other threshold regions. 
The method (QCD+NRQCD) consists in separating the $u,d,s$ contribution;  the 
 $\bar{c}c$ one is then treated as follows. 
If a resonance is below the channel for open charm production, which is set 
at $s=4m^2_c$ (with $c$ the pole mass of the $c$ quark), then it is considered  
as a bound state, and treated in the n.w.a. Above $\bar{c}c$ threshold, one uses nonrelativistic QCD.  
For our choice of $c$ quark mass, both $J/\psi$ and $\psi'$ should 
be considered to be below threshold.
 
The reasonable agreement, within errors, between the (QCD+NRQCD) result for the $\bar{c}c$ 
contributions in the  region
$3.7^2\;\gev^2\leq s\leq4.6^2\;\gev^2$, $10^{-5}\times (139\pm13)$ and the 
result obtained using experimental data only, $10^{-5}\times(112\pm6)$, 
gives one confidence to use the same theoretical method of calculation for the 
other thresholds where the quality of the experimental data is poorer, 
or these data are lacking. 
However, for the $\bar{c}c$ region we consider that the 
results based on experimental data are the best and thus write
$$-10^5\times \hat{\piv}^{(0)}_{\rm h}(M_Z^2;\;3.7^2\,\gev^2 \leq s\leq 4.6^2\,\gev^2)=
205\pm7.\equn{(3.7)}$$

\medskip
\noindent3.3.2. {\sl   $\bar{b}b$: $\upsilonv$, $\upsilonv'$ and the continuum
 $10.086^2\;\gev^2\leq s\leq 11.2^2\;\gev^2$}
\smallskip
\noindent 
For the region around $\bar{b}b$ threshold we can repeat the calculations as above. 
First, we add the contribution of the resonances below $\bar{b}b$ 
threshold, that we calculate in the n.w.a.:
$$\eqalign{
10^{-5}\times (5.8\pm 0.2)&\quad [\upsilonv]\cr
10^{-5}\times (2.1\pm0.1)&\quad [\upsilonv']\cr}$$
For the continuum we find, for a $b$ quark pole mass of\ref{10} $m_b=5.022\pm0.060\,\gev$,
$$\eqalign{
10^{-5}\times (57.9\pm0.1\,(\lambdav))&\quad [udsc];\;
\hbox{(QCD;  $10.086^2\;\gev^2\leq s\leq11.2^2\;\gev^2$)}\cr
10^{-5}\times (8.7\pm0.6\,(m_b))&\quad\hbox{[$\bar{b}b$]. (NRQCD)}\cr
\hbox{Sum}: 10^{-5}\times (66.6\pm0.6)}$$
\smallskip

If we had estimated the $\bar{b}b$ contribution 
saturating with the resonances $\upsilonv'',\;\dots\upsilonv^{\rm V}$, 
with electronic widths as given in ref.~14 
we would have got $5.2\pm1.2$ instead of the value 
$8.7\pm0.6$ that we found with the NRQCD calculation. 
We choose this last as our preferred value and write thus

$$-10^5\times \hat{\piv}^{(0)}_{\rm h}(M_Z^2;\;10.086^2\,\gev^2 \leq s\leq 11.2^2\,\gev^2)=
75\pm1.\equn{(3.8)}$$

\medskip
\noindent3.3.3. {\sl   $\bar{t}t$ threshold:  $T$ bound states and 
the continuum $348^2\;\gev^2\leq s\leq 360^2\;\gev^2$}
\smallskip
\noindent
The bound states produce a negligible contribution; for the ground state, a second order
 QCD calculation\ref{11} 
gives $\gammav(T\to e^+e^-)=12.5\pm1.5\,{\rm keV}$ and thus the contribution to 
$-10^5\times \hat{\piv}^{(0)}_{\rm h}$ is of $-0.11$. For the threshold region, 
a NRQCD calculation gives, for the $t$ quark contribution, $-0.47$, 
while the $udscb$ one is $-1.41$. (Note that in this calculation 
we are neglecting electroweak interactions, so 
we treat the $t$ quark as if it was stable).  All together, we find
$$-10^5\times \hat{\piv}^{(0)}_{\rm h}(M_Z^2; t\;{\rm thresh.})=-2.
\equn{(3.9)}$$
The error is negligible. 

The total contribution of the threshold regions is thus
$$-10^5\times \hat{\piv}^{(0)}_{\rm h}(M_Z^2; c,\,b,\,t\;\hbox{thresh's.})=278\pm7.
\equn{(3.10)}$$

\booksubsection{3.4.  The lowest order $\hat{\piv}^{(0)}_{\rm h}(M_Z^2)$}
Adding all the contributions to $\hat{\piv}^{(0)}_{\rm h}(M_Z^2)$ we get
$$10^5\times\deltav_{\rm had}^{(0)}\alpha(M_Z^2)=-10^5\times \hat{\piv}^{(0)}_{\rm h}(M_Z^2)=
2732\pm12,
\equn{(3.11)}$$  
or, excluding the top quark contribution,
$$10^5\times\deltav_{\rm had}^{(0)}\alpha^{(5)}(M_Z^2)=
2739\pm12.
\equn{(3.12)}$$

\booksection{4. The radiative corrections,  $-\hat{\piv}^{(1)}_{\rm h}$; 
the full  $-\hat{\piv}^{(0+1)}_{\rm h}$; $\bar{\alpha}_{\rm Q.E.D.}(M_Z^2)$}
\vskip-0.3cm
\booksubsection{4.1.  $-\hat{\piv}^{(1)}_{\rm h}$}
We have next the contribution of intermediate states containing a photon. 
At low energy ($s\leq 1.2\,\gev^2$) we evaluate them individually, and at high 
energy ($s\geq 1.2\,\gev^2$) with the parton model. For the second we have
a contribution equal to the 
zero order one (for which we take the result of the 
previous section) multiplied by the factor 
$$\dfrac{\sum_fQ^4_f}{\sum_fQ^2_f}\;\dfrac{3\alpha}{4\pi}.$$ 
This gives 
$$-10^5\times\hat{\piv}^{(1)}_{\rm h}(M_Z;\;s\geq 1.2\,\gev^2)=1.4\pm0.1,
\equn{(4.1)}$$
the error depending on what one does in the quark thresholds, 
especially around the narrow resonances ($J/\pi,\,\psi',\,\upsilonv,\,\upsilonv'$).
For the low energy region we repeat, with obvious changes, the analysis of \ty. 
Only the processes $\pi^+\pi^-\gamma$, $\pi^0\gamma$ and $\eta\gamma$ 
produce effects at the $10^{-5}$ level (respectively, $3.4\pm0.8$,  $2.9\pm0.2$ and $0.8\pm0.1$, 
in units of  $10^{-5}$). 
The first is evaluated in the narrow width approximation, or with a detailed 
calculation using theoretical formulas that relate $\pi\pi\gamma$ to $\pi\pi$, and 
taking into account experimental cuts; the details may be found in \ty. 
Both methods give essentially the same result. The other two are evaluated in the narrow width 
approximation, dominated by the $\rho\to\pi^0\gamma$, 
$\omega\to\pi^0\gamma$ and $\phi\to\eta\gamma$ 
contributions. 
In addition, the low energy  ($s<0.7^2\,\gev^2$)$\pi^0\gamma$ is calculated with a phenomenological coupling 
$\pi^0\gamma\gamma$, adjusted to reproduce the decay $\pi^0\to\gamma\gamma$. 
Again, the details are given in \ty. 

Adding all of this to (4.1) we find
$$-10^5\times\hat{\piv}^{(1)}_{\rm h}(M_Z^2)
=8.5\pm0.9.
\equn{(4.2)}$$

We summarize our results in Table~1:

\setbox0=\vbox{
\medskip
\setbox1=\vbox{\petit \offinterlineskip\hrule
\halign{
&\vrule#&\strut\hfil\quad#\quad\hfil&\vrule#&\strut\hfil\quad#\quad\hfil&
\vrule#&\strut\hfil\quad#\quad\hfil&\vrule#&\strut\hfil\ #\ \hfil\cr
 height2mm&\omit&&\omit&&\omit&&\omit&\cr 
&\hfil Channel\hfil&&\hfil Energy range\hfil&
&\hfil Method of calculation\hfil&
&\hfil ${{\rm Contribution\; to}}\atop{\textstyle -10^{5}\times \deltav_{\rm had}\alpha}$\hfil& \cr
height1mm&\omit&&\omit&&\omit&&\omit&\cr
\noalign{\hrule}height1mm&\omit&&\omit&&\omit&&\omit&\cr
&$\pi^+\pi^-$&&\vphantom{\Big|}$s\leq  0.8\,\gev^2$&
&\hfil Fit to $e^+e^-\,+\,\tau\,+\,{\rm spacel.}$ data\hfil&&
$307.6\pm3.6$& \cr
\noalign{\hrule}height1mm&\omit&&\omit&&\omit&&\omit&\cr
&\vphantom{\Big|}$\pi^+\pi^-$ 
\phantom{\big|}&&\phantom{\Big|}$0.8\leq s\leq  1.2\,\gev^2$&
&\hfil Fit to $e^+e^-$ data \hfil&&$27.3\pm0.6$& \cr
\noalign{\hrule}height1mm&\omit&&\omit&&\omit&&\omit&\cr
&$3\pi$&&\vphantom{\Big|}$s\leq  1.2\,\gev^2$&
&\hfil B.--W. + const. fit to $e^+e^-$  data\hfil&&$39.5\pm1.5$& \cr
\noalign{\hrule}height1mm&\omit&&\omit&&\omit&&\omit&\cr
&\vphantom{\Big|}$2K$&&$s\leq  1.2\,\gev^2$&
&\hfil B.--W. + const. fit to $e^+e^-$  data\hfil&&$41.6\pm1.3$& \cr
\noalign{\hrule}height1mm&\omit&&\omit&&\omit&&\omit&\cr
&\vphantom{\Big|}$4\pi,\;5\pi,\;\eta\pi\pi,\dots$&
&$s\leq  1.2\;\gev^2$&&\hfil Fit to $e^+e^-$  data \hfil&&$2.9\pm0.7$& \cr
\noalign{\hrule}height1mm&\omit&&\omit&&\omit&&\omit&\cr
&\vphantom{\Big|}Inclusive&&$1.2\,\gev^2\leq s\leq 2\,\gev^2$&
&\hfil  Fit to $e^+e^-$  data\hfil&&$53.1\pm5.3$& \cr
height1mm&\omit&&\omit&&\omit&&\omit&\cr
\noalign{\hrule}height1mm&\omit&&\omit&&\omit&&\omit&\cr
&\vphantom{\Big|}Inclusive; $uds$ &&$2\,\gev^2\leq s\leq  3.7^2\gev^2$&
&\hfil Perturbative QCD \hfil&&$330.2\pm2.4$& \cr
\noalign{\hrule}height1mm&\omit&&\omit&&\omit&&\omit&\cr
&\vphantom{\Big|}$J/\psi,\,\psi'$ && &&\hfil n.w.a. 
\hfil&&$93.5\pm5.0$&\cr
\noalign{\hrule}height1mm&\omit&&\omit&&\omit&&\omit&\cr
&\vphantom{\Big|}Inclusive&&$3.7^2\,\gev^2\leq s\leq  4.6^2\,\gev^2$&
&\hfil Fit to $e^+e^-$  data \hfil&&$111.8\pm5.5$& \cr
\noalign{\hrule}height1mm&\omit&&\omit&&\omit&&\omit&\cr
&\vphantom{\Big|}Inclusive; $udsc$ &&$4.6^2\,\gev^2\leq s\leq  10.086^2\,\gev^2$&
&\hfil  Perturbative QCD \hfil&&$421.3\pm0.8$& \cr
\noalign{\hrule}height1mm&\omit&&\omit&&\omit&&\omit&\cr
&\vphantom{\Big|}$\upsilonv,\,\upsilonv'$ && &&\hfil n.w.a. 
\hfil&&$7.9\pm0.2$&\cr
\noalign{\hrule}height1mm&\omit&&\omit&&\omit&&\omit&\cr
&\vphantom{\Big|}$b$ quark thresh.&&$10.086^2\,\gev^2\leq s\leq  11.2^2\,\gev^2$&
&\hfil Pert. + Nonrelativistic QCD \hfil&&$66.6\pm0.6$& \cr
\noalign{\hrule}height1mm&\omit&&\omit&&\omit&&\omit&\cr
&\vphantom{\Big|}Incl.; $udscb (t)$&&$11.2^2\,\gev^2\leq s\leq  \infty$ 
(except $t$ thr.)&
&\hfil Perturbative QCD \hfil&&$1\,230.3\pm3.4$& \cr
\noalign{\hrule}height1mm&\omit&&\omit&&\omit&&\omit&\cr
&\vphantom{\Big|}$t$ quark thresh.&&$348^2\,\gev^2\leq s\leq  360^2\,\gev^2$&
&\hfil Pert. + Nonrelativistic QCD \hfil&&$-2.0\pm0.1$& \cr
\noalign{\hrule}height1mm&\omit&&\omit&&\omit&&\omit&\cr
&\vphantom{\Big|}$\gamma+$ hadrons&&\hfil Full range\hfil&
&\hfil Various methods \hfil&&$8.5\pm0.9$& \cr
\noalign{\hrule}}
\vskip.05cm}
\centerline{\box1}
\smallskip
\centerline{\petit Table~1}
\centerrule{6cm}
\medskip
\noindent{Summary of contributions to $\deltav_{\rm had}\alpha=-\hat{\piv}_{\rm h}$.
 ``B.--W.
+ const." means a Breit--Wigner fit, including the correct  phase space factors, plus a
constant; note that only for the four narrow resonances
$J/\psi,\,\psi';\,\upsilonv,\,\upsilonv'$ 
we use the n.w.a. The errors are 
uncorrelated except those for QCD calculations 
(that have to be added linearly) and those for the $2\pi$ states (see text). 
 For the details of the final states $\gamma+$ hadrons
 we refer to \ty.}
\medskip}
\box0

\booksubsection{4.2. The full $\deltav_{\rm had}\alpha$; the QED coupling on the $Z$; 
discussion}
Adding then (4.2) to the lowest order expression we get the final result
$$10^5\times\deltav_{\rm had}\alpha(M_Z^2)=-10^5\times\left[\hat{\piv}^{(0)}_{\rm h}(M_Z^2)
+\hat{\piv}^{(1)}_{\rm h}(M_Z^2)
\right]=
2740\pm12,
\equn{(4.3)}$$
or, excluding the top quark contribution,
$$10^5\times\deltav_{\rm had}\alpha^{(5)}(M_Z^2)=
2747\pm12.
\equn{(4.4)}$$

The pure QED corrections amount to (see, e.g., ref.~15)
$$-10^5\times\hat{\piv}_{\rm QED}(M_Z^2)
=3149.7687.
\equn{(4.6)}$$
Adding this to (4.3) and using (1.1) we get the value for the running 
electromagnetic coupling
$$\bar{\alpha}_{\rm Q.E.D.}(M_Z^2)=
\dfrac{1}{128.965\pm0.017}.\equn{(4.7)}$$

When comparing with other determinations we will 
restrict ourselves to those performed {\sl after} the 
data from Novosibirsk\ref{3,4} and Beijing\ref{13} 
have become available; these experiments increase the 
set of data by almost one 
order of magnitude, and have much better precision than the older ones. 
Thus, one may consider older determinations as superseded. 
So we compare our results with the
 determinations of refs.~15,16. 
We have,
$$10^5\times\deltav_{\rm had}\alpha^{(5)}(M_Z^2)=
\cases{
2743\pm19\;/\;2765\pm21,\quad \hbox{(MOR)}\cr
2761\pm36,\quad \hbox{(BP)}\cr
2790\pm40,\quad \hbox{(J)}\cr}
\equn{(4.5)}$$
In the MOR determination, the two values depend on the method of 
calculation used.
As a general rule, comparing any two of the results quoted above, the largest
reason for a reduced error bar is related to a wider use of perturbative
QCD. For instance, the analysis of MOR uses perturbative QCD in the regions
$2.8\,\gev^2 \leq s\leq 3.7^2\,\gev^2$, and  $s\geq 5^2\,\gev^2$.
We have used pertubative QCD in the region 
$2\,\gev^2 \leq s\leq 3.7^2\,\gev^2$, justified in view of its agreement 
with the precise new experimental data
(as discussed in the text), and the regions above the $\bar{c}c$ and 
$\bar{b}b$ thresholds. Here, the use of the calculations incorporating the 
exact effect of the quark masses has allowed to get a precise determination
for $4.6^2\,\gev^2 \leq s \leq 5^2\,\gev^2$, where the experimental data are not
very precise, but (again) perfectly consistent with QCD.

In conclusion, we have performed a detailed evaluation of the hadronic contributions
to the running electromagnetic coupling, obtaining a substantially reduced error bar.
The ingredients are the following:
First, we use Novosibirsk ($e^+ e^-$) and LEP ($\tau$) data to fit the $2\pi$ contribution.
Invoking the analyticity and unitarity properties of the pion form factor
allows to include spacelike data also, improving the compatibility of the $e^+ e^-$ data 
with the results from $\tau$ decay, and reducing the corresponding error.
Second, the low energy $3\pi$ and $2K$ states have been considered individually, 
after the latest Novosibirsk data on the $\omega$ and $\phi$ resonances. We perform a full-fledged
fit, including the exact threshold factors.
Third, we have used perturbative QCD in the region $s \geq 2\,\gev^2$ (away from quark thresholds).
In particular, the recent LEP $\tau\to\nu_\tau+{\rm hadrons}$, 
and BES $e^+e^-\to{\rm hadrons}$ data justify the QCD result for 
$2\,\gev^2 \leq s\leq 3^2\,\gev^2$, implying the largest part of the error reduction.
We also use the Beijing\ref{13} data, thus gaining precision, for the contribution in the energy 
range $3.7^2\,\gev^2$ to $4.6^2\,\gev^2$. Last but not least, the next order radiative corrections have 
been taken into account. This is essential for our calculation as the radiative contribution
is indeed of the same order of the final error bar.

\vfill\eject

We thank M.~Gr\"unewald for valuable discussions.
The financial support of CICYT is  gratefully acknowledged.

\booksection{References}
{\petit
\item{1.-}{\/{\sc J. F. de Troc\'oniz and F. J. Yndur\'ain}, FTUAM 01-08 
(hep-ph/0106025).}
\item{2.-}{For analyticity 
properties, and for $\pi\pi$ scattering, see {\sc B. R. Martin, D. Morgan and G.~Shaw}, {\sl 
Pion-Pion Interactions in particle Physics}, Academic Press, New York 1976.
 This text also gives information on the 
Omn\`es-Muskhelishvili method for which more details may be found in
{\sc N.~I.~Muskhelishvili}, {\sl Singular
 Integral Equations}, Nordhoof, 1958. The effective range theory can be seen 
very clearly explained
 in {\sc H.~Pilkuhn}, {\sl The interaction of 
hadrons}, North-Holland, Amsterdam, 1967.}
\item{3.-}{\/$e^+e^-$ data, $\rho$ region:\prajnyp{L.~M.~Barkov et al.}{Nucl. Phys.}{B256}{1985}{365}; 
 {\sc R. R. Akhmetsin et al.} Budker INP 99-10 (hep-ex/9904027). 
$\tau$ decay data, Aleph: \prajnyp{R. Barate et al.}{Z.Phys.}{C76}{1997}{15},
{\sl Eur.Phys.J.} {\bf C4} (1998) 409; 
Opal: \prajnyp{K.~Ackerstaff et al.}{Eur.Phys.J.}{C7}{1999}{571}. Data on $F_\pi(t)$ at spacelike $t$: 
(NA7)  \prajnyp{S.~R.~Amendolia et al.}{Nucl. Phys.}{B277}{1986}{168}.} 
\item{4.-}{\/$e^+e^-$ data, $\omega$ and $\phi$ region, $KK$ and $3\pi$:
  \prajnyp{R. R. Akhmetshin et al.,}{Phys. Letters}{B466}{1999}{385 and 392}; 
ibid., {\bf B434} (1998) 426 and ibid., {\bf B476} (2000) 33. 
 \prajnyp{M.~N.~Achasov et al.}{Phys. Rev.}{D63}{2001}{072002};   
{\sl Phys. Letters} {\bf B462} (1999) 365 and   
 Preprint Budker INP 98-65, 1998 (hep-ex/9809013). 
$\phi\to2\pi$:  \prajnyp{M.~N.~Achasov et al.}{Phys. Letters}{B474}{2000}{188}.}
\item{5.-}{$KK$\/: \prajnyp{P. M. Ivanov et al.}{Phys. Letters}{107B}{1981}{297}; $3\pi$: 
\prajnyp{A.~Cordier et al.}{Nucl. Phys.}{B172}{1980}{13}; $4\pi$ and more: 
\prajnyp{G.~Cosme et al.}{Nucl. Phys.}{B152}{1979}{215}. For a review, see  
\prajnyp{S. Dolinsky et al.}{Phys. Reports}{C202}{1991}{99}, and work quoted there. 
All these references 
give results for energies below $t=2\,\gev^2$.}
\item{6.-}{\prajnyp{K. G. Chetyrkin, A. L. Kataev and F. V. Tkachov}{Phys. Letters}{B85}{1979}{277}; 
\prajnyp{M.~Dine and J.~Sapiristein}{Phys. Rev.  Letters}{43}{1979}{277};
\prajnyp{W.~Celmaster  and R.~Gonsalves}{Phys. Rev.  Letters}{44}{1979}{560} 
($O(\alpha_s^2$). \prajnyp{S.~G. Gorishny, A. L. Kataev and S.~A.~Larin}{Phys. Letters}{B259}{1991}{144} 
($O(\alpha_s^3$). For the $O(m^2)$, $O(m^4)$ corrections, 
see \prajnyp{K.~G.~Chetyrkin and J.~H.~K\"uhn}{Phys. Letters}{B248}{1990}{359};
 \prajnyp{E.~Braaten, S.~Narison and A.~Pich}{Nucl. Phys.}{B373}{1992}{581} 
 and work quoted there.}
\item{7.-}{\prajnyp{K. Adel and F.~J.~Yndur\'ain}{Phys. Rev.}{92}{1998}{113}; 
\prajnyp{Adel and F.~J.~Yndur\'ain}{Rev. Acad. Ciencias (Esp.)}{92}{1998}{113} 
(hep-ph/9509378). 
For a clear exposition of NRQCD calculations see \prajnyp{N.~Brambilla, A.~Pineda, 
J.~Soto and A.~Vairo}{Nucl. Phys.}{B566}{2000}{275}.}
\item{8.-}{\/{\sc J.~Schwinger}, {\sl Particles, Sources and Fields}, Vol.~2. Addison-Wesley, 1973.}
\item{9.-}{\prajnyp{S. Bethke}{J. Phys.}{G26}{2000}{R27}; 
{\sc J.~Santiago and F. J. Yndur\'ain},  FTUAM 01-01, 2001 (hep-ph/0102247);
{\sc D. Strom}, ``Electroweak measurements on the $Z$ resonance", 
Talk presented at the 5th Int. Symposium on Radiative Corrections, RadCor2000, 
Carmel, Ca., September 2000.}
\item{10.-}{\prajnyp{S.~Titard and 
F.~J.~Yndur\'ain}{Phys. Rev.}{D49}{1994}{6007}, to one loop;
 \prajnyp{A. Pineda and F.~J.~Yndur\'ain}{Phys. Rev.}{D58}{1998}{094022} 
and ibid., {\bf D61}, 077505 (2000), to two loops.}
\item{11.-}{\/{\sc F. J. Yndur\'ain}, hep-ph/0007333, to be published in
 the Proc. E. Schr\"odinger Institute Meeting on Confinement. Vienna, 2000.}
\item{12.-}{\prajnyp{B.~Abbott et al.}{Phys. Rev.}{D58}{1998}{052001}; 
\prajnyp{T.~Affolder et al.}{Phys. Rev.}{D64}{2001}{032003}.}
\item{13.-}{BES: {\sc J. Z. Bai et al}, hep-ex/0102003.}
\item{14.-}{\prajnyp{D. Besson et al. (CLEO)}{Phys. Rev. Letters}{54}{1985}{381}; 
Particle Data Group: \prajnyp{D. E. Groom et al.}{Eur. Phys. J.}{C15}{2000}{1}.}
\item{15.-}{(EJ): \prajnyp{S. Eidelman and F.~Jegerlehner}{Z. Phys.}{C67}{1995}{585}; 
updated in (J): {\sc F.~Jegerlehner,} DESY 01-028 (hep-ph/0104304), 
 Symposium in honor of A.~Sirlin, New York University, 2000.}
\item{16.-}{\/(MOR): \prajnyp{A.~D. Martin, 
J. Outhwaite and M. G. Ryskin}{Phys. Letters}{B492}{2000}{69}, 
{\sl Eur. Phys. J.} {\bf C19} (2001) 681;
(BP): {\sc H.~Burkhardt and B.~Pietrzyk}, LAPP-EXP 2001-03 (2001).}
\item{}{}
}

\bye